\documentstyle[12pt]{article}
\begin{document}
\tolerance=5000
\def\be{\begin{equation}}
\def\ee{\end{equation}}
\def\bea{\begin{eqnarray}}
\def\eea{\end{eqnarray}}
\def\nn{\nonumber \\}
\def\cF{{\cal F}}
\def\det{{\rm det\,}}
\def\Tr{{\rm Tr\,}}
\def\e{{\rm e}}
\def\etal{{\it et al.}}
\def\erp2{{\rm e}^{2\rho}}
\def\erm2{{\rm e}^{-2\rho}}
\def\er4{{\rm e}^{4\rho}}
\def\etal{{\it et al.}}
\def\gsim{\ ^>\llap{$_\sim$}\ }

\ 

\vskip -2cm

\ \hfill
\begin{minipage}{3.5cm}
NDA-FP-53 \\
December 1998 \\
\end{minipage}

\vfill

\begin{center}
{\Large\bf  Two-Boundaries AdS/CFT Correspondence 
in Dilatonic Gravity}

\vfill

{\sc Shin'ichi NOJIRI}\footnote{\scriptsize 
e-mail: nojiri@cc.nda.ac.jp} and
{\sc Sergei D. ODINTSOV$^{\spadesuit}$}\footnote{\scriptsize 
e-mail: odintsov@mail.tomsknet.ru}

\vfill

{\sl Department of Mathematics and Physics \\
National Defence Academy, 
Hashirimizu Yokosuka 239, JAPAN}

\ 

{\sl $\spadesuit$ 
Tomsk Pedagogical University, 634041 Tomsk, RUSSIA \\
}

\ 

\vfill

{\bf abstract}

\end{center}

We discuss dilatonic gravity (bulk theory) from the point of 
view of (generalized) AdS/CFT correspondence. 
Self-consistent dilatonic background is considered. It 
may be understood as two boundaries space where AdS boundary 
appears as infinite boundary and new (singular) boundary 
occurs at short distances. The two-point correlation 
function and conformal dimension for minimal and dilaton 
coupled scalar are found. Even for minimal scalar, the  
conformal dimension is found to be different on above two 
boundaries. Hence, new CFT appears in AdS/CFT correspondence 
at short distances. AdS/CFT correspondence may be understood 
as interpolating bulk theory between two different CFTs.

\newpage

AdS/CFT correspondence \cite{Mal,W,GKP} gives an interesting 
framework to relate (classical) bulk theory with the 
conformal field theory living on the infinite boundary. 
In the original version of AdS/CFT corespondence, it has only 
one boundary (AdS). However, there are some indications 
\cite{BST} that singularities which appear and in bulk and 
in boundary theories could mean the opening of a new space. 
For example, it happens \cite{BST} that singular branes 
become regular in a dual, conformally rescaled, frame. 
That may indicate that there naturally appears second boundary 
(vacuum) in AdS/CFT correspndence \cite{BST,Behrndt}. 
In other words, brane probably should not reside only at the 
infinite boundary of AdS \cite{Behrndt} and it is better to 
imagine the brane is everywhere.

In the present letter, we discuss AdS/CFT correspondence for 
dilatonic theories in the space with two boundaries. We 
concentrate on the behaviour of correlation function of 
scalar field in different cases (minimal scalars, dilaton 
coupled scalars (see \cite{NO} for introduction)) on 
new short distance boundary. It is shown that CFT on this 
boundary  is different from the one on 
infinite AdS boundary. The shift of conformal dimension due 
to mass and curvature coupling is also calculated. 
Finally, we give some remarks about another representation 
of dilatonic gravity under discussion as higher derivative 
gravity.

We start 
from the following action of dilatonic gravity
 in $d+1$ dimensions:
\be
\label{i}
S=-\int d^{d+1}x \sqrt{-g}\left(R - \Lambda 
- \alpha g^{\mu\nu}\partial_\mu \phi \partial_\nu \phi \right)\ .
\ee
In the following, we assume $\lambda^2\equiv -\Lambda$ 
and $\alpha$ to be positive.
From the variation of the metric $g^{\mu\nu}$, we obtain\footnote{
The conventions of curvatures are given by
\begin{eqnarray*}
R&=&g^{\mu\nu}R_{\mu\nu} \\
R_{\mu\nu}&=& -\Gamma^\lambda_{\mu\lambda,\kappa}
+ \Gamma^\lambda_{\mu\kappa,\lambda}
- \Gamma^\eta_{\mu\lambda}\Gamma^\lambda_{\kappa\eta}
+ \Gamma^\eta_{\mu\kappa}\Gamma^\lambda_{\lambda\eta} \\
\Gamma^\eta_{\mu\lambda}&=&{1 \over 2}g^{\eta\nu}\left(
g_{\mu\nu,\lambda} + g_{\lambda\nu,\mu} - g_{\mu\lambda,\nu} 
\right)\ .
\end{eqnarray*}
}
\be
\label{ii}
0=R_{\mu\nu}-{1 \over 2}g_{\mu\nu}R + {\Lambda \over 2}g_{\mu\nu}
+ \alpha \left(\partial_\mu\phi\partial_\nu\phi 
-{1 \over 2}g_{\mu\nu}g^{\rho\sigma}\partial_\rho \phi
\partial_\sigma \phi \right)
\ee
and from that of dilaton $\phi$
\be
\label{iii}
0=\partial_\mu\left(\sqrt{-g}g^{\mu\nu}\partial_\nu\phi\right)\ .
\ee
We assume that solutions for $g_{\mu\nu}$ and $\phi$ depend 
only on one of the coordinate, say $y\equiv x^d$
\be
\label{iv}
g_{\mu\nu}=g_{\mu\nu}(y)\ ,\ \ \phi=\phi(y)
\ee
and $g_{\mu\nu}$ has the following form
\be
\label{v}
ds^2=\sum_{\mu,\nu=0}^d g_{\mu\nu}dx^\mu dx^\nu
=f(y)dy^2 + g(y)\sum_{i,j=0}^{d-1}\eta_{ij}dx^i dx^j
\ee
Here $\eta_{ij}$ is the metric in the flat (Lorentzian) background.
Then the equations of motion (\ref{ii}) and (\ref{iii}) take the 
following forms:
\bea
\label{vi}
0&=&-{d(d-1) \over 8}\left({g' \over g}\right)^2 + {\lambda^2 \over 2}f 
+ {\alpha \over 2}(\phi')^2 \\
\label{vii}
0&=&-{d-1 \over 2}{g'' \over g} + {d-1 \over 4}{f'g' \over fg}
-{(d-1)(d-4) \over 8}\left({g' \over g}\right)^2 
+ {\lambda^2 \over 2}f - {\alpha \over 2}(\phi')^2 \\
\label{viii}
0&=&\left(\sqrt{{g^d \over f}}\phi'\right)'\ .
\eea
Here $'$ expresses the derivative with respect to $y$. Eq.(\ref{vi}) 
corresponds to $(\mu,\nu)=(d,d)$ in (\ref{ii}) and Eq.(\ref{vii}) to 
$(\mu,\nu)=(i,j)$. The case of $(\mu,\nu)=(0,i)$ or $(i,0)$ is identically 
satisfied.
 Integrating (\ref{viii}), we find 
\be
\label{ix}
\phi'=c\sqrt{{f \over g^d}}\ .
\ee
 Substituting (\ref{ix}) into (\ref{vi}), we can solve it algebraically 
with respect to $f$:
\be
\label{x}
f={d(d-1) (g')^2 \over 4g^2\left(
\lambda^2 + {\alpha c^2 \over g^d}\right)}\ .
\ee
We find that Eq.(\ref{vii}) is automatically satisfied when we substitute 
(\ref{ix}) and (\ref{x}). Therefore $g$ can be an arbitrary function of $y$ 
but this corresponds to the degree of the freedom of the reparametrization 
of $y$ in the metric (\ref{v}). We can fix it by choosing 
\be
\label{xi}
g=y\ .
\ee
Then we find from (\ref{ix}) and (\ref{x}), 
\bea
\label{xii}
f&=&{d(d-1) \over 4y^2\left(
\lambda^2 + {\alpha c^2 \over y^d}\right)} \\
\label{xiii}
\phi&=&c\int dy \sqrt{{d(d-1) \over 
4y^{d+2}\left(\lambda^2 + {\alpha c^2 \over y^d}\right)}} \nn
&=&\phi_0+{1 \over 2}\sqrt{(d-1) \over d\alpha}\ln\left\{
{2\alpha c^2 \over \lambda^2 y^d}+1 \pm\sqrt{
\left({2\alpha c^2 \over \lambda^2 y^d}+1\right)^2 -1}\right\}\ .
\eea
 From Eq.(\ref{xiii}), we find the dilaton field behaves 
as $\phi\rightarrow \mp\sqrt{d(d-1) \over \alpha}\ln y^d$
when $y\rightarrow 0$.  
Since the string coupling constant $\e^\phi$ should be small in order that 
the supergravity picture is consistent, the dilaton field cannot grow 
up positively. Therefore the $-$ sign should be chosen in the $\pm$ sign 
in Eq.(\ref{xiii}) and we find
\be
\label{xiiia}
\phi\rightarrow {1 \over 2}\sqrt{d(d-1) \over \alpha}\ln y\ .
\ee
The metric given by (\ref{v}), (\ref{xi}) and (\ref{xii}) becomes 
that of the usual anti-de Sitter space in the limit where $\alpha$ or 
$c$ vanishes.

The backreaction from the non-trivial background of dilaton (\ref{xiii}) 
to the metric as in (\ref{xii}) changes the structure of the spacetime 
near the region where $y=0$. When $y$ is small, the metric behaves as
\be
\label{xiv}
ds^2\sim {d(d-1) \over 4\alpha c^2}y^{d-2}dy^2 
+ y\sum_{i,j=0}^{d-1}\eta_{ij}dx^i dx^j\ .
\ee
The metric tells that the distance $l$ between the point 
with finite $y=y_0$ and that of $y=0$ is finite:
\be
\label{xv}
l=\int_{y=y_0}^{y=0}ds = \int_0^{y_0} dy 
\left({d(d-1) \over 4\alpha c^2}\right)^{1 \over 2}y^{d-2 \over 2}
=\left({(d-1) \over 4d\alpha c^2}\right)^{1 \over 2}y_0^{d \over 2}\ .
\ee
This should be compared with the case of usual anti-de Sitter space where 
the distance is infinite. In the limit where $\alpha$ or $c$ vanishes, 
the distance $l$ in (\ref{xv}) becomes infinite as expected.
It ahould be also noted that there is a curvature singularity at $y=0$ 
since the scalar curvature is given by 
\be
\label{xxxviib}
R = -{(d+1) \Lambda \over d-1} - \alpha c^2 y^{-d}
\sim \alpha c^2 y^{-d} \ .
\ee

The infinite boundary discussed in AdS/CFT correspondence lies at $y=\infty$. 
When $y\rightarrow \infty$, $f$ and $\phi$ in the solution (\ref{xii}), 
(\ref{xiii}) behave as 
\be
\label{ci}
f\rightarrow {d(d-1) \over 4\lambda^2 y^2}\left(1 +{\cal O}(y^{-d})
\right)\ ,\ \   
\phi\rightarrow \phi_0\left(1 + {\cal O}(y^{-{d \over 2}})\right)\ .
\ee
Therefore the geometry of spacetime asymptotically approaches to 
that of AdS, which tells the correlation functions of matter fields 
on the boundary corresponding to $y=\infty$ do not change with those 
on the boundary of AdS. 

After Wick rotating the spacetime signature 
by changing $x^0\rightarrow ix^0$, as an example, we consider free 
massless scalar whose action is given by
\be
\label{xvi}
S^\chi ={1 \over 2}\int d^{d+1}x \sqrt{g}g^{\mu\nu}\partial_\mu\chi
\partial_\nu\chi
\ee
and consider the correlation function in the neighborhood of the boundary 
 $y=\infty$ by solving the following equation
\bea
\label{cii}
\sqrt{g}\Box G(y, X^2)&\sim& A_0 \partial_y\left(
y^{{d \over 2}+1}\partial_y G(y, X^2)\right)
+ {y^{{d \over 2}-2} \over A_0}\sum_{i=0}^{d-1}\partial_i^2
G(y, X^2) \nn
&=&0 \ , \\
\sqrt{g}\Box &\equiv&\partial_\mu\left(\sqrt{g}g^{\mu\nu}
\partial_\nu\right)\ , \nn
A_0&\equiv& 2\sqrt{\lambda^2 \over d(d-1)} \nn
X^2&\equiv&\sum_{i=0}^{d-1} (x_1^i - x_2^i)^2 \ .
\nonumber 
\eea
The solution of above equation near the infinite boundary  is given by
\be
\label{ciii}
G(y, X^2) = G_0 y^{-{d \over 2}} 
\left\{\left({4 \over A_0^2}y^{-1} +  X^2 
\right)^{-d} + {\cal O}(y^{-d})\right\}\ .
\ee
Here $G_0$ is a constant. Equation (\ref{ciii}) gives a correlation 
function at $y=\infty$ as in \cite{W}. In the limit of
$y\rightarrow \infty$, we obtain
\be
\label{cix}
G(y=0, X^2) = G_0 y^{-{d \over 2}}\left( X^2 \right)^{-d}\ .
\ee
This correlation function is nothing but that of the operators 
with the conformal dimension $d$ in some kind of conformal 
field theory.

Since there is a singularity at $y=0$, we need to check if the fields 
at the boundary $y=\infty$ have a unique extension in the bulk spacetime 
$y<\infty$. As an example, we consider free massless scalar in (\ref{xvi}). 
Since $\sqrt{g}$ is given by 
\be
\label{vx}
\sqrt{g}=\sqrt{d(d-1) y^{d-2}\over 4\left(
\lambda^2 + {\alpha c^2 \over y^d}\right)} \ ,
\ee
the square-integrability requires 
\bea
\label{vxi}
&\chi \sim o(y^{-{d \over 4}})\ &\mbox{when}\ y\rightarrow \infty \nn
&\chi \sim o(y^{-{d \over 2}})\ &\mbox{when}\ y\rightarrow 0\ .
\eea
If the Laplace equation $\Box\chi=0$ has 
a square integrable solution whose boundary value at $y=\infty$ vanishes, 
the uniqueness is broken since we can add the solution to any given 
solution. The solution can be written using the Fourier transformation 
with respect to the coordinates $\{x^i,\ i=1,\cdots,d\}$ as follows:
\be
\label{vxii}
\chi(y,x^i)={1 \over (2\pi)^{d \over 2}}\int d^dk f_k(y)
\e^{i\sum_{i=0}^{d-1} k_ix^i}\ .
\ee
Then the Laplace equation is rewritten as
\bea
\label{vxiib}
0&=&\sqrt{g}\Box \chi \nn
&\rightarrow&{2 \over \sqrt{d(d-1)}}\partial_y\left(y^{{d \over 2}+1}
\sqrt{\lambda^2+{\alpha c^2 \over y^d}}\partial_y f_k(y)\right)
\nn 
&& - {\sqrt{d(d-1)} \over 2}{k^2 y^{{d \over 2}-1} \over 
\sqrt{\lambda^2+{\alpha c^2 \over y^d}}}f_k(y)\ .
\eea
 Multiplying $f_k^*(y)$ (the complex conjugate of $f_k(y)$) and 
integrating with respect to $y$, we obtain
\bea
\label{vxiii}
0&=&\int_0^\infty dy \left\{{2y^{{d \over 2}+1} \over \sqrt{d(d-1)}}
\sqrt{\lambda^2+{\alpha c^2 \over y^d}}\left|
\partial_y f_k(y)\right|^2
+ {\sqrt{d(d-1)} \over 2}{k^2 y^{{d \over 2}-1} \over 
\sqrt{\lambda^2+{\alpha c^2 \over y^d}}}\left|
f_k(y)\right|^2\right\} \nn
&& - \left.{2\sqrt{\alpha c^2 \over d(d-1)}}y
f^*_k(y)
\partial_y f_k(y)\right|_{y\rightarrow 0}\ .
\eea
Here we assume Eq.(\ref{vxi}) and that $\chi$ vanishes at the 
boundary $y=\infty$ ($f_k(\infty)=0$). Eq.(\ref{vxiii}) tells 
that if $f_k(y)$ does not vanish at $y=0$\footnote{
More exactly, if $f(y) \sim \sqrt{-\ln y}$ when $y\sim 0$, 
the boundary term in (\ref{vxiii}) becomes finite. The more 
singular behaviour of $f_k(y)$ can be consistent with the conditon 
of the square integrability in (\ref{vxi}). In such a case, the boundary 
term, and therefore the bulk integration, in (\ref{vxiii}) diverges.
}, there can be non-trivial square-integrable solutions whose boundary 
value at $y=\infty$ vanishes. This situation is very different from that 
in the usual AdS and the boundary value $\chi(y=\infty,x^i)$ 
cannot uniquely determine the value of $\chi$ in the bulk
$y<\infty$. Eq.(\ref{vxiii}) tells, however, 
$f_k(y)=\partial_yf_k(y)=0$ everywhere if $\chi$ vanishes at 
$y=0$ ($f_k(0)=0$) and there is no any non-trivial 
square-integrable solution. Note that the square-integrability requires 
$\chi$ vanish at $y=\infty$ due to (\ref{vxi}).
This implies that the value 
$\chi(y=0,x^i)$ at $y=0$ can uniquely determine the value of $\chi$ 
in the bulk $y>0$. Since there is a curvature singularity at $y=0$
(\ref{xxxviib}), $y=0$ can be regarded as a boundary, which is similar 
to the case of Schwarzschild spacetime although the singularity 
discussed here is naked\footnote{If $\alpha<0$, there is a horizon since 
$f(y)$ in (\ref{xii}) diverges at $y^d=-{\alpha c^2 \over \lambda^2}$). 
In this case, the singularity is not naked.
}. From the metric in (\ref{v}), the new boundary 
at $y=0$ has also the topology of $d$-dimensional Minkowski space in 
the Minkowski signature. 
In the usual AdS/CFT correspondence, the boundary at $y=\infty$ 
can be regarded as a brane in superstring or M-theory. Since the 
solution (\ref{xiii}) tells that there is a source of dilaton at 
$y=0$, the object at $y=0$ could be considered as brane with 
a dilatonic hair. (For classification of non-singular branes see \cite{GHT}).
If it is so, it is not so unnatural to expect that some kind of 
conformal field theory is realized on the boundary at $y=0$(like it 
happened in the example of 3d AdS gravity in ref.\cite{Behrndt}).

We now consider the correlation function of free massless scalar whose 
action is given by (\ref{xvi}).
We only consider the neighborhood of the boundary $y\sim 0$ and solve 
the following equation for correlation function 
\bea
\label{xvii}
\sqrt{g}\Box G(y, X^2)&\sim& A \partial_y\left(y\partial_y G(y, X^2)\right)
+ {y^{d-2} \over A}\sum_{i=0}^{d-1}\partial_i^2
G(y, X^2) \nn
&=&0 \ , \\
\sqrt{g}\Box &\equiv&\partial_\mu\left(\sqrt{g}g^{\mu\nu}
\partial_\nu\right)\ , \nn
A&\equiv& 2\sqrt{\alpha c^2 \over d(d-1)} \nn
X^2&\equiv&\sum_{i=0}^{d-1} (x_1^i - x_2^i)^2 \ .
\nonumber 
\eea
The solution of (\ref{xvii}) near the boundary $y\sim \infty$ is given by
\be
\label{xviii}
G(y, X^2) = G_0 \left({4 \over (d-1)^2 A^2}y^{d-1} +  X^2 
\right)^{-{d \over 2}}\ .
\ee
Here $G_0$ is a constant. Equation (\ref{xviii}) would give a correlation 
function on the boundary \cite{W}. In fact, in the limit of
$y\rightarrow 0$, 
we obtain
\be
\label{xix}
G(y=0, X^2) = G_0 \left( X^2 \right)^{-{d \over 2}}\ .
\ee
The correlation function (\ref{xix}) (which is different from one
on the infinite boundary) is nothing but that of the operators 
with the conformal dimension ${d \over 2}$ in some kind of conformal 
field theory. Especially when $d=2$, the correlation function is that of 
the product of left-moving and right-moving free fermions ${\cal O}
(x^1,x^2)\equiv\psi(z)
\psi^*(z^*)$ ($z=x^1+ix^2$).

As a more general case, we consider the correlation function of 
massless dilaton coupled scalar  whose 
action is given by
\be
\label{xx}
S^\chi_2 ={1 \over 2}\int d^{d+1}x \sqrt{g}
\e^{2\beta \sqrt{\alpha \over d(d-1)}\left(\phi-\phi_0\right)} 
g^{\mu\nu}\partial_\mu\chi \partial_\nu\chi\ .
\ee
Here $\beta$ is a parameter which is now introduced. 
From (\ref{xiiia}), we find near the boundary
\be
\label{xxi}
\e^{2\beta \sqrt{\alpha \over d(d-1)}\left(\phi-\phi_0\right)} 
\sim y^\beta\ .
\ee
Then in order to find the correlation function in the neighborhood of 
the boundary $y\sim 0$, we should solve  
the following equation instead of (\ref{xvii}):
\bea
\label{xxii}
\sqrt{g}\Box^\phi G^\phi(y, X^2)
&\sim& A \partial_y\left(y^{\beta+1}\partial_y G^\phi(y, X^2)\right)
+ {y^{\beta+d-1} \over A}\sum_{i=0}^{d-1}\partial_i^2 G^\phi(y, X^2) \nn
&=&0 \ , \\
\sqrt{g}\Box^\phi &\equiv&\partial_\mu\left(
\e^{2\beta \sqrt{\alpha \over d(d-1)} \left(\phi-\phi_0\right)} 
\sqrt{g}g^{\mu\nu} \partial_\nu\right)\ . \nonumber
\eea
The solution of (\ref{xxii}) near the boundary $y\sim 0$ is given by
\be
\label{xxiii}
G^\phi(y, X^2) = G_0 \left({4 \over (d-1)^2A^2}y^{d-1} + X^2 
\right)^{-{d \over 2}-{\beta \over d-1}}\ .
\ee
In the limit of $y\rightarrow 0$, we obtain
\be
\label{xxiv}
G^\phi(y=0, X^2) = G_0 \left( X^2 \right)^{-{d \over 2}
-{\beta \over d-1}}\ .
\ee
The correlation function (\ref{xxiv}) is that of the operators 
with the conformal dimension ${d \over 2}+{\beta \over d-1}$.
It is interesting that the conformal dimension is shifted by the 
parameter $\beta$ which comes from the dilaton coupling in (\ref{xx}). 
In the usual AdS/CFT correspondence, such a shift comes from the mass 
of the scalar field \cite{W}.

It would be also interesting to investigate the effect from the mass term.
We add the following dilaton dependent mass term to  
 the action (\ref{xx}):
\be
\label{xxxii}
S^m =-{m^2 \over 2}\int d^{d+1}x \sqrt{g}
\e^{2\gamma \sqrt{\alpha \over d(d-1)}\left(\phi-\phi_0\right)} 
\chi^2\ .
\ee
Then the equation coresponding to (\ref{xvii}) is given by
\bea
\label{xxxiii}
&& \sqrt{g}\left(\Box^\phi - m^2 
\e^{2\gamma \sqrt{\alpha \over d(d-1)}\left(\phi-\phi_0\right)} 
\right)G^m(y, X^2) \nn
&& \sim A \partial_y\left(y^{\beta+1}\partial_y G^m(y, X^2)\right)
+ {y^{\beta+d-1} \over A}\sum_{i=0}^{d-1}\partial_i^2
G^m(y, X^2) \nn 
&& - {m^2 \over A} y^{\gamma + d -3} G^m(y, X^2) \nn
&&=0 \ .
\eea
In order that $G^m(y, X^2)$ corresponds to the correlation function
of the conformal field theory on the boundary, $G^m(y, X^2)$ should have 
the following form in the limit of $y\rightarrow 0$:
\be
\label{xxxiv}
G^m \sim {y^\rho \over \left(X^2\right)^\nu}\ .
\ee
 Substituting (\ref{xxxiv}) into (\ref{xxxiii}), we find that $\gamma$ in 
(\ref{xxxii}) cannot be arbitrary but $\gamma$ should be given by
\be
\label{xxxv}
\gamma=\beta - d + 2\ .
\ee
When the equation (\ref{xxxv}) is satisfied, solution of (\ref{xxxii}) 
near the boundary $y\sim 0$ is given by
\be
\label{xxxvii}
G^m(y, X^2) = G_0 y^{-\beta\pm\sqrt{\beta^2+{4m^2 \over A^2}} \over 2}
\left({4 \over (d-1)^2A^2}y^{d-1} + X^2 
\right)^{-{d \over 2}\mp{ \sqrt{\beta^2+{4m^2 \over A^2}} \over d-1}}\ .
\ee
In the limit of $y\rightarrow 0$, we obtain
\be
\label{xxxvb}
G^m(y=0, X^2) = G_0 y^{-\beta\pm\sqrt{\beta^2+{4m^2 \over A^2}} 
\over 2}\left( X^2 \right)^{-{d \over 2}
\mp{ \sqrt{\beta^2+{4m^2 \over A^2}} \over d-1}}\ .
\ee
The correlation function (\ref{xxxvb}) is that of the operators 
with the conformal dimension ${d \over 2}
\pm {\sqrt{\beta^2+{4m^2 \over A^2}} \over d-1}$.

We can also consider the coupling of the matter scalar field $\chi$ with 
the scalar curvature $R$\cite{NO}:
\be
\label{xxxvib}
S^R =-{\mu^2 \over 2}\int d^{d+1}x \sqrt{g}R
\e^{2\delta \sqrt{\alpha \over d(d-1)}\left(\phi-\phi_0\right)} 
\chi^2\ .
\ee
Near the boundary $y\sim 0$ the behavior of $R$ is given by (\ref{xxxviib}). 
 Comparing (\ref{xxxviib}) with (\ref{xxi}), we can identify 
near the boundary
\be
\label{xxxviii}
R \sim \alpha c^2 \e^{-2\sqrt{d\alpha \over (d-1)}\left(\phi-\phi_0\right)} 
\ .
\ee
Therefore by the following replacement,
\be
\label{xxxix}
\gamma \rightarrow \delta - d \ ,\ \ \ 
m^2 \rightarrow \mu^2 \alpha c^2
\ee
we can use the results in (\ref{xxxv}), (\ref{xxxvii}) and (\ref{xxxvb}).

Finally we show that the gravity theory in (\ref{i}) can be rewritten as a 
higher derivative gravity theory without dilaton. When we rescale 
the metric by 
\be
\label{xxv}
g_{\mu\nu}\rightarrow \e^\rho g_{\mu\nu}\ ,
\ee
the action (\ref{i}) is rewritten after the partial integration as 
follows
\be
\label{xxvi}
S=-\int d^{d+1}x \sqrt{-g}\e^{{d-1 \over 2}\rho}\left(R - 
\Lambda\e^\rho + {(d-1)d \over 4}
g^{\mu\nu}\partial_\mu \rho \partial_\nu \rho
- \alpha 
g^{\mu\nu}\partial_\mu \phi \partial_\nu \phi \right)\ .
\ee
 Choosing $\rho$ as 
\be
\label{xxvii}
\rho=2\sqrt{\alpha \over d(d-1)}\phi
\ee
we obtain
\be
\label{xxviib}
S=-\int d^{d+1}x \sqrt{-g}\left\{\Phi^{d-1}
\left(R - \Lambda\Phi^2\right)\right\}\ .
\ee
Here
\be
\label{xxviii}
\Phi\equiv\e^{-\sqrt{\alpha \over d(d-1)}\phi}\ .
\ee
By using the equation of motion with respect to $\Phi$, 
we can solve 
$\Phi$ with respect to the scalar curvature $R$:
\be
\label{xxix}
\Phi^2={d-1 \over (d+1) \Lambda}R\ .
\ee
 Substituting (\ref{xxix}) into (\ref{xxviib}), we obtain 
the higher derivative gravity theory which contains 
${d+1 \over 2}$-power of the scalar curvature $R$:
\be
\label{xxx}
S=-{2 \over d+1}\left(-{d-1 \over (d+1)\Lambda}\right)^{d-1}\int
d^{d+1}x \sqrt{-g} (-R)^{d+1 \over 2}\ .
\ee
In the usual AdS/CFT correspondence, we believe that the conformal 
symmetry on the boundary manifold comes from the $SO(d,2)$ symmetry 
in the anti-de Sitter space but the spacetime given by (\ref{v}), 
(\ref{xi}) and (\ref{xii}) has no the $SO(d,2)$ symmetry.
The obtained correlation functions (\ref{xix}), (\ref{xxiv}) and 
(\ref{xxxvb}), however, seem to be those in some kind of conformal field 
theory. We should note that the action (\ref{xxx}) is invariant under 
the global scale transformation with a constant parameter $c$ 
\be
\label{xxxx}
g_{\mu\nu}\rightarrow \e^c g_{\mu\nu}\ ,
\ee
which might be the origin of the conformal symmetry on the boundary.

In summary, we discussed CFT which appears on (short 
distance) $y=0$ boundary of dilatonic  spacetime under consideration 
in generalized AdS/CFT correspondence. 
Dilatonic gravity may be considered as bulk theory 
interpolating between two different CFTs living at the 
boundaries $y=0$ and $y=\infty$. It could be really 
interesting to study AdS/CFT correspondence for the 
spacetimes with second boundary (of above sort) in N=4
conformal supergravity \cite{KTN,BRW,FT} (more exactly 
gauged supergravity) where at $y=\infty$ boundary 
${\cal N}=4$ super Yang-Mills theory lies \cite{LT} as 
dilaton naturally appears there.

\ 

\noindent
{\bf Acknowledgements.}  We would like to acknowledge 
helpful discussions with T. Hollowood, P.van Nieuwenhuizen 
and E. Mottola. 
The work by SDO has been partially supported by T-8, LANL 
and RFBR project n96-02-16017.

\end{document}